# Atmosphere Pressure Chemical Vapor Deposition of Graphene

Phuong V. Pham

SKKU Advanced Institute of Nano Technology (SAINT), Sungkyunkwan University (SKKU), Suwon, Gyeonggi-do 440-746, Republic of Korea. Email: pvphuong@skku.edu

Center for Multidimensional Carbon Materials, Institute for Basic Science, 44919, Ulsan, Republic of Korea.

## Abstract

Recently, graphene is of highly interest owing to its outstanding conductivity, mechanical strength, thermal stability, etc. Among various graphene synthesis methods, atmosphere pressure chemical vapor deposition (APCVD) is one of the best synthesis one due to very low diffusitivity coefficient and a critical step for graphene-based device fabrication. High-temperature APCVD processes for thin film productions are being recognized in many diversity technologies such as solid-state electronic devices, in particular, high quality epitaxial semiconductor films for silicon bipolar and metal oxide semiconductor (MOS) transistors. Graphene-based devices exhibit high potential for applications in flexible electronics, optoelectronics, and energy harvesting. In this chapter, recent advances of APCVD-based graphene synthesis and their related applications will be addressed.

## 1. Introduction

Single-layer graphene (SLG), bilayer graphene (BLG), and multi-layer graphene (MLG) films have been regarded as optimal materials for electronics and optoelectronics owing to their excellent electrical properties and their ability to integrate with current top-down device fabrication technology [1-99]. Since the beginning of the 21$^{st}$ century, the interest in graphene materials has drastically increased, which is apparent in the number of annual publications on graphene (**Figure 1**). Until now, various strategies, including chemical vapour deposition (CVD) [28], liquid and mechanical exfoliation from graphite [23,29,30], epitaxial growth on crystal substrate [31-34], or solution-based processes on graphene oxides (GO) [35-41]. They have been investigated for obtaining graphene layers. In particular, recent advances in CVD growth have successfully led to large-scale graphene production on metal substrates [1,27,42-48], driven by the high demand for utilizing graphene in possible applications of current complementary metal-oxide-semiconductor (CMOS) technology such as radio-frequency transistors, optical devices, and deposition processes [2].

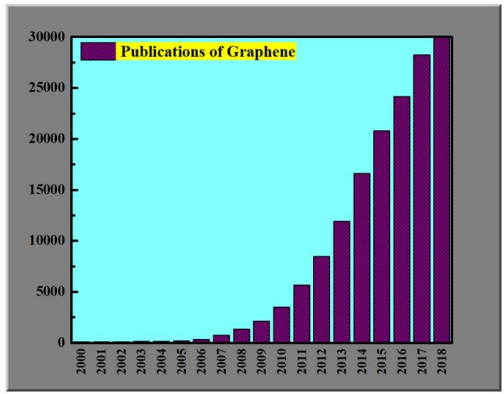

**Figure 1.** Graphene Publications from 2000 to 2018. Source: ISI Web of Science (search: Topic = Graphene).

High-quality large-scale graphene has been synthesized on conducting metallic substrates by using the catalytic CVD growth approach, which promoted a wide range of graphene-based device applications [1,28,42-49]. However, graphene grown on a metal substrate needs to be transferred onto dielectric substrates for electronic applications. Although various approaches, such as wet etching/transfer [45], mechanical exfoliation/transfer [23,29,30], bubbling transfer [50], electrochemical delamination [51-53], for transferring from the catalytic metal growth substrates to dielectric device substrates have been developed, none of these approaches is free from degradation of the transferred graphene. For example, 'wet etching and transfer', the most widely used tranfer approach, is a serial process, which includes encapsulation the graphene surface with polymer supporting.

The growing through CVD is for production the films or for coating of metal, semiconductor, crystalline and vitreous formed-compounds in either, occupying high-purity as well as desirable characteristics. In addtion, the creating of controllable film with varying stoichiometry resulting CVD uniquely among deposition approaches. Other advantages include reasonably low-cost of the equipment and operating expenses, suitability for semicontinuous operation. Consequently, the variants of CVD have been developed recently such as low-pressure chemical vapor deposition (LPCVD), APCVD, plasma-enhanced chemical vapor deposition (PECVD), and laser-enhanced chemical vapor deposition (LECVD). The hybrid system represents for both CVD and physical vapor deposition (PVD) have also discovered.

Among the utilized synthesis methods, APCVD has been considered as the most potential and medium-cost one for large-scale high-quality graphene on various metallic substrates including Pt [54], Ir [55], Ni [56], and Cu [57]. Especially, Cu is considered as the best choice owing to low-carbon solubility, well-controlled surface, and inexpensive for growing monolayer graphene [57,58]. Many efforts have made for obtaining large-scale single crystal graphene with as less grain boundaries as possible. There are two general methods to realize: the first method associates with the growth of single domains with possibly enlargement [54,59-66]. Despite the centimeter-scale domains have achieved, this method is no the bright candidate in practice for large-scale growth because of the difficult quantity control of nucleation seeds as well as unclear self-limiting growth factors resulting worse and requiring very long growth time (over 24 h) [67,68]. The second method is the alignment of graphene domain orientations on arbitrary substrates and then to atomically stitching them to form uniform single crystalline graphene [69,70]. This seems to be ideally for growing large-scale.

The CVD method creates large-scale graphene but polycrystalline morphology, which includes different oriented-domains. Such the orientated disorders inevitably lead to graphene grain boundaries (GGBs) formation at intercalated interface of domain.[71-73] GGBs include a series of non-hexagonal rings of pentagons, heptagons, and octagons. Intrinsic graphene has high conductivity and chemical inert, however, the apprearance of atomically defect lines and GGBs in graphene could remarkably modulate its features such as mechanics,[73,74] electrics,[75,76] chemistry,[77,78] and magnets.[79] These defect lines have extraordinary features and relies the atomical configuration at GGBs and the crystallinity of mono-grains, moreover, they could be modulated through different function groups.[79] Therefore, the investigation on the orientations and boundaries of the grains is the key to comprehend the underlying features as well as to realize compatible applications of graphene.

Conventionally, graphene is grown on metal foils (Cu, Ni, Pt, etc.) through APCVD, LPCVD, or PECVD. The temperatures, pressures and concentrations of precursor gases inside the furnace play the key roles on the graphene growth and quality. These factors need be optimized to get the desirable growth results. In general, surface morphology is studied via optical microscope (OM) and scanning electron microscopy (SEM); the graphene quality is examined through Raman spectra, UV-visible spectroscopy, and transmission electron microscopy (TEM); and sheet resistance ($R_s$) of synthesized graphene is measured utilizing four-point probe technique. Almost graphene films synthesized via APCVD are monolayer with better quality in atmospheric condition compared with LPCVD or PECVD. The synthesis study using APCVD would be greatly significant in the desirable growth of graphene and other related materials. In this chapter, recent advances of APCVD-based graphene synthesis with practically concerns about chemical vapor transport, deposition process as well as their device applications will be addressed.

## 2. Growth Mechanism of APCVD-based Graphene

APCVD growth of graphene is a chemical process at atmosphere pressure for the formation of SLG or FLG on an arbitrary substrate by exposing the substrate to the gas-phase precursors at controlled reaction conditions.[80] Owing to the versatile nature of APCVD, intricately mixed homogeneous gas-phase and heterogeneous surface reactions are involved [81]. In general, as the partial pressure and/or temperature in the reaction substances are increased, homogeneous gas-phase reactions and the resulting homogeneous nucleation become significant [81]. To grow a high-quality graphene layer, this homogeneous nucleation needs be minimized [81]. A general mechanism for APCVD-based graphene growth on catalytic metal substrates, for the growth of uniform and highly crystalline graphene layer on the surface, includes eight steps as follow: (1) mass transport of the reactant, (2) reaction of the film precursor, (3) diffusion of gas molecules, (4) adsorption of the precursor, (5) diffusion of the precursor into substrate, (6) surface reaction, (7) desorption of the product, and (8) removal of the by-product (**Figure 2**) [82].

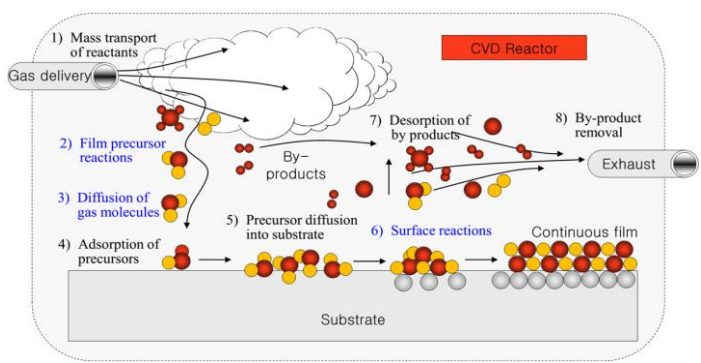

**Figure 2.** Diagram of CVD growth mechanism (APCVD and LPCVD) of graphene: transport and reaction processes. Reproduced with permission from [82]. Copyright 2011, Freund Publishing.

It has been a general observation in most of the experimental works that the LPCVD produces non-uniformly thick graphene layers. However, as high quality single layer graphene growth, then APCVD would be a better choice due to reduced mean free path/diffusion coefficient of the reactive species on the catalyst. [83] APCVD shows a much lower diffusitivity coeficient: $D_g \sim 1/$(total pressure) compare with LPCVD. As the result, less small multilayer graphene islands appearing on a full large monolayer graphene island via APCVD growth process.[83]

In another graphene growth mechanism model, hydrocarbon molecules are absorbed as well as dissociated on Cu forming active carbon species by dehydrogenation reaction (**Figure 3**). These species are diffused on two sides of Cu foil and agglomerated on its active sites to form graphene nucleation seeds. Actually, introducing $H_2$ gas is mandatory as key role in graphene growth for most CVD approaches. The overall processes of graphene growth on Cu is described (**Figure 3**).[93] Generally, there are three main expected steps: i) adsorp-decompose, ii) diffuse-desorb, and iii) nucleate-grow (**Figure 3**).[93] The active carbon species are commonly no stable and easily agglomerate with thermodynamically stable species on active sites to form graphene nucleation seeds by the reactions: $(CH_x)_s$ + graphene → (graphene-C) + $(CH_x)_s$ with x= 1,2,3.[93] Till those seeds formed, almost the active carbon species are incorporated and captured at surface/interface of graphene lattice. In addition, $H_2$ precursor plays more role as an etchant and controls the size as well as shape of graphene domains via reactions: $H_s$ + graphene → (graphene-C) + $(CH_x)_s$ with x = 1,2,3.[93]

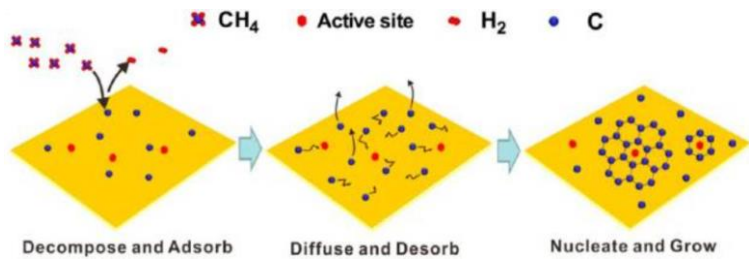

**Figure 3.** Schematic for graphene growth mechanism on Cu substrate. Reproduced with permission from [93]. Copyright 2012, American Chemical Society.

Typically, APCVD growth of 2D materials (e.g. graphene) involves catalytic activation of chemical reactions of precursors at the growth substrate surface/interface in a properly designed environment. Generally speaking, the roles of precursors, conditions (e.g. fast growth rates, large domain size, or very high crystalline quality), atmosphere, substrates and catalysts are the key factors affecting the final quality of the grown 2D materials. So far, significant efforts have been made to prepare highly crystalline 2D materials (e.g. graphene), but many challenges are still ahead. For example, due to the rough feature of catalytic metal surface, growth of uniform and high quality graphene is considerably difficult. The 2D material research community is also interested in new precursors (e.g. solid precursor only, gas precursor or solid precursor mixed with certain solvents) that could form uniformly high-quality graphene with minimal defect density. Another question is the effect of growth rate on the catalytic metal surface on the quality of graphene. Currently, it is difficult to give an exact answer, as investigations are progressing at an exponential rate. To date, the understanding of the concept of the general mechanism of the APCVD growth of graphene is still not yet adequate, neither experimentally nor theoretically. Thus, understanding the graphene growth mechanism and the effect of various growth conditions will be of significant interest to the 2D material research community to obtain large-scale, high-quality graphene.

## 3. APCVD Growth of Graphene

CVD is a thin solid film deposition process of vapor species through suitably chemical reaction. The deposition needs low-carbon solubility substrate in high temperature region [84]. To date, CVD is the sole approach which could produce high-quality graphene with ultra-large size [85,45]. For the first experiment, researchers successfully synthesized graphene films via CVD on Ni and Cu catalytic substrates [44,86,87]. The significant progress for large-size and high quality graphene films have also well-done [85,87].

APCVD requires high temperature (~1000 °C) for graphene synthesis. It makes expensive experimentally, therefore, it requires the CVD equipments more sophisticatedly.

Unfortunately, this is an obstacle for direct-growth of graphene on insulating device substrate (e.g., $SiO_2$), for instance, it can produce unavoidable physical damages around 1000 °C and dramatically degrades the qualify of synthesized graphene. At the result, the deposition of graphene on insulating substrates at reduced temperatures becomes very necessary [88,89]. The features of graphene will change corresponding with the number of layers. Hence, the electronic, optical, mechanical and other features of graphene could be tuned by controlling the number of layers, or by adjusting the experimental conditions. Speakingly, it is a huge challenge to exactly atomic-scale control the number of graphene layers [90-92]. The proposed synthesis method is compound by four stages (cleaning, precursor injection, reaction time, and cooling). The temperature and fowrate of carbon precursor were the studied parameters. In **Figure 4**a-b, the common CVD systems and diferences are shown. [94]

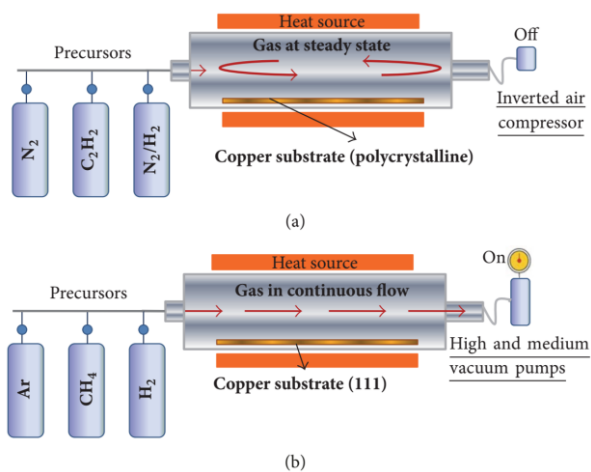

**Figure 4.** Schematic of graphene synthesis by (a) APCVD using polycrystalline Cu substrate and gas discontinuous fow and (b) high vacuum using single crystalline Cu(111) substrate and gas continuous fow. Reproduced with permission from [94]. Copyright 2018, Hindawi Publishing.

In 2016, Wang et al obviously showed millimeter-scale graphene single crystals synthesized on Cu through APCVD (**Figure 5**).[95] The possible mechanism will base on graphene nucleation and growth kinetic. At the stage of nucleation, thermal decomposition of oxide layer leads to $O_2$ desorption at high temperature at front side of Cu and dominates the temperature dependence of nucleation density. The graphene island growth is edge-limited on two sides of Cu at various enlargement rates. The roughness of support substrates (quartz, sapphire) also affects the graphene deposition. After optimized annealing and polished support substrate, the isolated graphene islands (~3 mm) were produced with a growth rate (25 μm/min). The domains were uniformly single-crystalline graphene with good mobility (~4900 cm$^2$/V.s) at room temperature. **Figure 5**b-f shows OM images of graphene domains/Cu for 60 min growth at various Ar pre-annealing times. The growth and cooling parameters maintained the same. Consequently, the Ar heating induced the growth of individual graphene domains with relatively low nucleation density (102 cm$^{-2}$) owing to the catalyst passivation of oxide layer on top of Cu. By introducing the following annealing process to 30 min (**Figure 5**d), the nucleation density continuously decreased to ~12 nuclei/cm$^2$, and a grain size of ~1 mm was obtained after 60 min growth (**Figure 5**e). However, the extending of annealing (120 min) caused the increase of nucleation density with reduced domain size (~0.6 mm) (**Figure 5**f). The eolutions of nucleation densitys and domain sizes via annealing are shown in **Figure 5**g. The nucleation density reveals a non-monotonic dependence of Ar-passivated annealing. On Cu surface with the oxide layer, a competing mechanism should be taken in the annealing at growth temperature. Since the surface becomes flatter after annealing, the O-species also gradually desorb at surface which leaves pristine Cu as active nucleation sites following the growth steps.

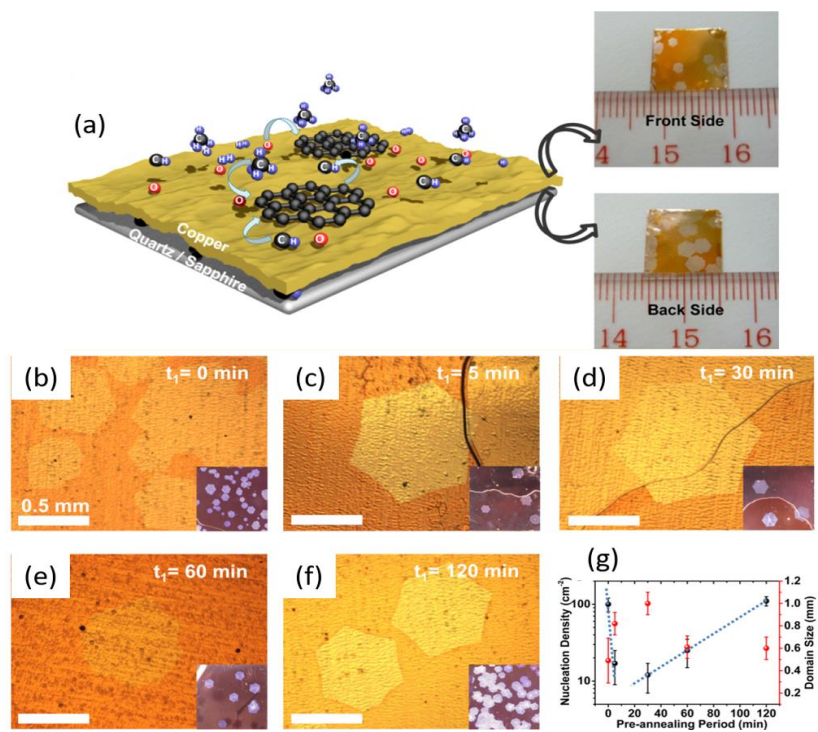

**Figure 5.** (a) Schematic for APCVD-grown graphene on Cu located on quartz or sapphire substrates. Insets in (a) are OM images of graphene/Cu at front side and back side. OM images of grown graphene on Cu foils with various pre-annealing times of (b) 0, (c) 5, (d) 30, (e) 60, (f) 120 min at the same growth time (60 min). (g) Evolution of nucleation density and size of graphene domains as a function of pre-annealing times. Reproduced with permission from [95]. Copyright 2016, American Chemical Society.

Thermodynamics of CVD-based graphene synthesis using catalytic Cu at particular temperature is the same irrespective at atmosphere pressure (AP), low pressure (LP), or ultrahigh vacuum (UHV).[83] But, the kinetics will vary on the different steps. The kinetics of cooling rate, the pressure of CVD furnace has major ramification on the graphene growth rate, large-scale thickness uniformity, and the defect density. **Figure 6**A illustrates a steady state flow of a mixture of $CH_4$, $H_2$, and Ar precursors on Cu surface at ~1000 °C.[83] The boundary layer because of steady state gas flow is stagnant. The carbon species firstly (1) diffuse via the boundary layer and reach the surface, and at the surface they get (2) adsorbed on the surface, (3) decompose for formation of active carbon species, (4) diffuse on/into the catalyst surface and forming the graphene lattice, (5) inactive species (e.g. $H_2$) get desorbed from the surface, forming $H_2$, and (6) diffuse away from the surface through boundary layer and are finally swept away by the bulk gas flow.[83] Processes which occur on/close the surface are highly affected by substrate temperature. Generally, there are two fluxes of active species that co-exist: flux of active species via boundary layer and the rate at which the active species are consumed at metal substrate surface forming graphene lattice (**Figure 6**B).[83] The equations of these fluxes are $F_{mass\ transport} = h_g(C_g - C_s)$ and $F_{surface\ reaction} = K_s C_s$ where, $F_{mass\text{-}transport}$ is the flux of active species through the boundary layer, $F_{surface\text{-}reaction}$ is the flux of consumed active species at surface, $h_g$ is the mass transport coefficient, $K_s$ is the surface reaction constant, $C_g$ is the concentration of gas in bulk, and $C_s$ is the concentration of active species at the surface.[83] At high temperatures, under APCVD parameters, mass transport via boundary layer is rate limiting ($K_s \gg h_g$), and under LP and UHV parameters, the surface reaction is the rate limiting step ($h_g \gg K_s$).[83]

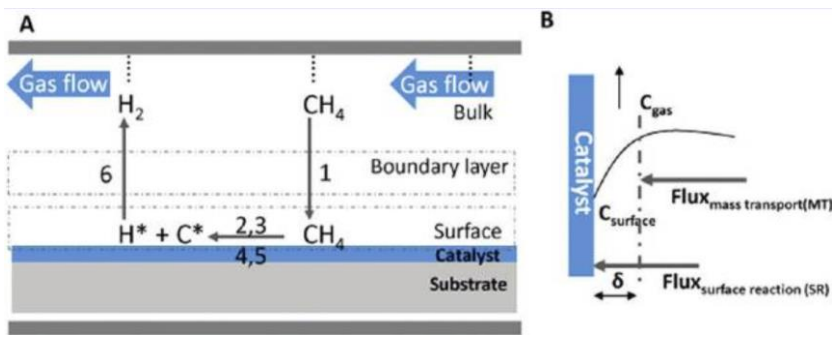

**Figure 6.** (A) Schematic for graphene growth mechanism using low carbon solid solubility catalysts (Cu) at atmosphere pressure (AP), low pressure (LP), or ultrahigh vacuum (UHV) environments. (B) Mass transport and surface reaction fluxes under steady state conditions. Reproduced with permission from [83]. Copyright 2010, American Chemical Society.

A diagram of designed-APCVD for graphene synthesis is described in **Figure 7**. This splits one gas inlet in two paths by means of glass valves: one for heating, annealing and cooling (P1); and the other for graphene synthesis (P2). The synthesis temperature is 980-990 °C in tubular furnace; during heating, the Ar-$H_2$ (5% of $H_2$) flow is 0.2 l/min via P1. Then Cu is annealed in 20-30 min. Then, the flow passed via alcohol container (P2) to desired growth rate, and the varied time depended on precursor type. Keeping the flow rate, the flux is to P1 and the sample is annealed in 10 min. As the result, this annealing step improved the graphene quality a lot. The temperature, the time, and flow rate permit a sophisticated controlling on carbon concentration toward tubular furnace (**Figure 7**) [96].

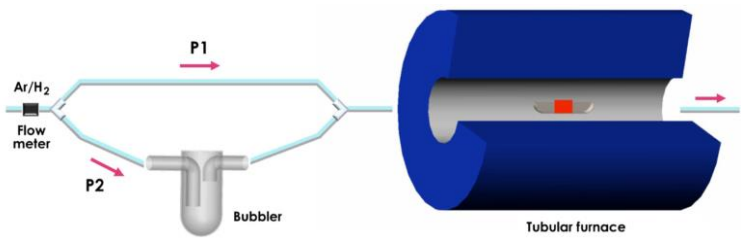

**Figure 7.** Graphene growth mechanism on new designed APCVD with a bubbler. A single gas inlet is split into two paths (P1 and P2) using glass. During heating/annealing/cooling, gas is passed through P1; while for graphene growth, gas flows through P2 and bubbler cotaining alcohol. Inside the tubular furnace a combustion boat is placed in the center containing a Cu foil. Reproduced with permission from [96]. Copyright 2013, Elsevier.

The additional oxide substrate ($SiO_2$, $Al_2O_3$) to continuously supply the oxygen ($O_2$) to Cu surface of APCVD graphene synthesis (**Figure 8**a,b).[97] The Cu was placed on oxide substrates away 15 μm gap. Last report proved that the oxide shall slowly release $O_2$ at above 800 °C.[100] To clear it, Xu et al carried out Auger electron spectroscopy (AES) element measurement on $SiO_2$, as the result, $O_2$ escaped from $SiO_2$ after annealing in UHV or CVD (**Figure 8**c).[97] Despite the small amount of $O_2$ released, the $O_2$ concentration between the narrow gap (15 μm) of Cu and oxide substrate could high because of the trapping effect, hence the $O_2$ attachment to Cu surface significantly trapped. Beside AP condition, high $CH_4$ flow (5 sccm) and high $CH_4/H_2$ ratio (~1) ensure a sufficient supply of carbon for ultrafast domain synthesis. The large domains (~0.3 mm) appeared on the back surface of Cu (**Figure 8**d). Contrarily, graphene domains at front surface of Cu are 20 times smaller (~15 μm) (**Figure 8**e). All these domains are star-shaped. For further demonstration, when graphite is used as the supporting substrate, graphene domains on both sides of Cu are also star-shaped similar with the case of $SiO_2$ (**Figure 8**g).[97]

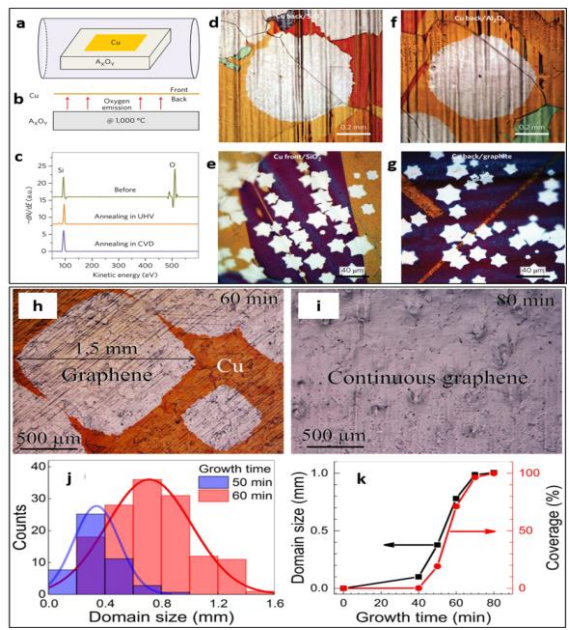

**Figure 8.** a) Schematic of graphene growth on Cu using $O_2$-assisted APCVD. b) Side view of (a). c) Auger electron spectroscopy (AES) of $SiO_2$ before/after annealing at 1000 °C in UHV and APCVD systems. The disappearance of $O_2$ peak after annealing proves the emission of $O_2$ from the oxide at high temperature. d,e, OM images of graphene domains on the back (d) and front (e) Cu surfaces using fused silica as the supporting substrate. f,g, OM images of graphene domains on the back Cu surfaces using sapphire ($Al_2O_3$) (f) or graphite (g). Reproduced with permission from [97]. Copyright 2016, Nature Publishing Group. (h,i) OM images of graphene/Cu after 60 and 80 min, respectively. APCVD conditions: 1 atm, 1030 °C, 0.3 sccm $CH_4$, 80 sccm $H_2$, and 3920 sccm Ar. (j) Histogram of domain size for 50 and 60 min growth. (k) The graphene coverage and the average domain size as a function of growth time. Reproduced with permission from [98]. Copyright 2016, Nature Publishing Group.

Figure 8h,i revealed OM images of graphene/Cu in 60 and 80 min under $CH_4$ 0.000075 Pa.[98] The graphenes needed to place in air (200 °C, 1 min) for chemical oxidation between Cu and air to emerge the nonoxidized-graphene domains for the observation by OM and human-eye. The graphene partly covered the Cu (~70%) after 60 min. Some isolated domains were ~1.5 mm (**Figure 8**h).[98] But the color as well as contrast of sample were no change afer 80 min implying that Cu got full coverage by graphene via OM image (**Figure 8**i).[98] The histogram in **Figure 8**j showed the growth distribution of domain sizes for 50-60 min.[98] The distribution of domain sizes increased via growth time; the domain sizes were 0.4-0.8 mm for 50-60 min. The time evolution of average domain size and coverage were clearly described in **Figure 8**k.[98] The domain sizes for 100% coverage after 80 min was ~1 mm.[98]

A growth at APCVD in Mohsin et al is to avoid Cu evaporation from the foil substrate which commonly appears in LPCVD case.[99,101] Regarding the **Figure 9,** firstly Cu is placed on a tungsten (W) foil for preventing the dewetting of liquid Cu on quartz. Then, it was heated up 1100 $^0$C which is higher melting point of Cu for 30 min under Ar (940 sccm) and H$_2$ (60 sccm). Then, the temperature is lowered to 1075 $^0$C, and the Cu resolidified. Growth was carried out at this temperature, with 0.1% dilute CH$_4$ in Ar. On the right side of **Figure 9**, it shows OM image of two hexagon domains of graphene (>1 mm) which has similar morphology.[99] These domains obtained in Mohsin et al are hexagons with very less roughness edges compared to previous reports.[63,93]

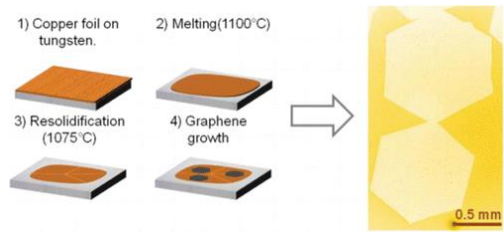

**Figure 9.** (a) Simple method to grow millimeter-size graphene single crystals on melted and resolidified Cu using APCVD. (b) OM image of the synthesized graphene domains. Reproduced with permission from [99]. Copyright 2013, American Chemical Society.

APCVD-based bilayer graphene growth is another great issue for optimum electronic and photonic devices because of higher carrier mobility and wider band gap by perpendicular electric field compared with single-layer graphene.[102,103] The synthesis of bilayer graphene faced many drawbacks owing to limited grain size and nonsynchronic growth between the first and the second graphene layer. In 2016, Sun et al reported a cooling-APCVD to growth of bilayer graphene on polycrystalline Cu (**Figure 10**).[102] Here, the surface adsorption for decomposed carbons and phase segregation for dissolved carbon were shown. Consequently, the milimeter-scale hexagonal bilayer graphene (~1.0 mm) was produced. This study opens the possibility to grow centimeter-scale bilayer graphene for optimum graphene-based applications in recent future.

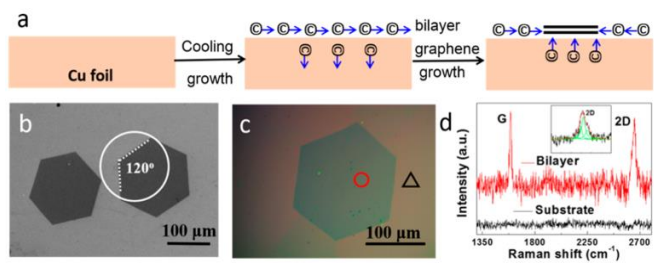

**Figure 10.** (a) Schematic of bilayer graphene growth using a cooling APCVD. (b,c) SEM and OM images of bilayer graphene domains, respectively. (d) Rama spectra of bilayer graphene. Reproduced with permission from [102]. Copyright 2016, American Chemical Society.

## 4. Conclusions

Strategies for direct graphene growth using APCVD method have been reviewed. The prospects of APCVD-grown graphene are bright and currently receiving considerable attention from the 2D material research community. Yet, understanding the growth process and conditions that affect the quality of graphene is not adequate.

Because APCVD growth depends on thermal decomposition of carbon resource, the growth rate is usually low and size of the graphene domain is small, resulting in growth of defective graphene layer. To date, the challenge remains for this research directions and large-scale high-quality graphene productions are still hard. In order to obtain more advanced results, an in-depth mechanism understanding of APCVD-based graphene growth is essential.

The Cu is considered as the best substrate owing to low-carbon solubility, well-controlled surface, and inexpensive for growing monolayer graphene. Generally, graphene is synthesized on Cu via APCVD paramaters utilizing $CH_4$ precursor revealed that the growth is different from single layer graphene at low $CH_4$ concentration to multi-layer domain on a single layer at higher $CH_4$ concentration. In particular, APCVD-based graphene growth at higher $CH_4$ concentration has no self-limiting compared with LPCVD, indicating that further investigation is needed to point out the growth mechanism. The kinetic processes utilizing low carbon solubility catalyst via APCVD parameters were presented.

Finally, APCVD-grown graphene on flexible/metal/dielectric substrates assisted by metal powder precursors (solid, gas, solution) contained inside a sub-chamber for direct evaporation into an innovative APCVD main-chamber allowing hexagonal domain formation on flexible/metal/dielectric substrates is the secret topic and needs to be exploited in the coming time.

## Conflict of Interest

There are no conflicts of interest to declare.